\documentclass[11pt,reprint,showpacs,showkeys,amsfonts,amssymb,amsmath,nofootinbib]{article}
\usepackage{bm,amsmath,epsf,epsfig,latexsym,amssymb,cite,lmodern}
\usepackage{graphics,psfrag,empheq,hyperref,float}
\def\Xint#1{\mathchoice
   {\XXint\displaystyle\textstyle{#1}}%
   {\XXint\textstyle\scriptstyle{#1}}%
   {\XXint\scriptstyle\scriptscriptstyle{#1}}%
   {\XXint\scriptscriptstyle\scriptscriptstyle{#1}}%
   \!\int}
\def\XXint#1#2#3{{\setbox0=\hbox{$#1{#2#3}{\int}$}
     \vcenter{\hbox{$#2#3$}}\kern-.5\wd0}}

\def\dashint{\Xint-}

\newcommand{\lm}{\lambda}

\newcommand{\cl}{{\cal L}}

\newcommand{\vp}{{\mathbf{p}}}

\newcommand{\vq}{{\mathbf{q}}}

\newcommand{\vr}{\mathbf{r}}

\newcommand{\bg}{\begin{align}}
\newcommand{\eeg}{\end{align}}
\newcommand{\be}{\begin{equation}}
\newcommand{\ee}{\end{equation}}
\newcommand{\ba}{\begin{eqnarray}}
\newcommand{\ea}{\end{eqnarray}}

\newcommand{\nn}{\nonumber}

\newcommand{\ep}{\epsilon}

\newcommand{\email}[1]{\footnote{{\em } \texttt{#1}}}

\begin{document}

\thispagestyle{empty}

\title{Unitarizing non-relativistic  Coulomb scattering}
\author{J.~A. Oller\email{oller@um.es}   \\[0.5em]
{\it \small Departamento de F\'{\i}sica. Universidad de Murcia. E-30071,
Murcia. Spain.}
} 

\maketitle
\begin{abstract}
  We compare the exactly solvable nonrelativistic Coulomb scattering with two recent unitarization methods for infinite-range forces.
  These methods require to calculate perturbatively the corresponding partial-wave amplitudes, which are then unitarized.  
  We calculate the Coulomb partial-wave amplitudes up to the one-loop order.
  On the one hand, the unitarization method  developed by 
  Refs.~\cite{Blas:2020och,Blas:2020dyg} reproduces properly the exact solution, with an accuracy improving as the order in the perturbative calculation of the input perturbative partial-wave amplitudes  increases. This is also shown to be the case for the pole position of the ground state. 
  On the other hand, the method developed by 
  the more recent Ref.~\cite{Dobado:2022} gives rise to partial-wave amplitudes 
  that do not reproduce the known solvable solution, and gives rise to a pole position with zero binding energy.
\end{abstract}

\tableofcontents

\section{Introduction}

Recently, Ref.~\cite{Blas:2020och} has developed a method to unitarize infinite-range interactions in partial-wave amplitudes (PWAs)
from their perturbative calculation. In this way non-perturbative PWAs result, which fulfill unitarity exactly and account for the right-hand or unitarity cut (RC). By the same token these unitarized amplitudes give rise to poles. In particular, Ref.~\cite{Blas:2020och} applied this approach to study graviton-graviton scattering and predicted the existence of a relatively light resonance (the graviball) with quantum vacuum numbers, $J^{PC}=0^{++}$, at $s_P=(0.07-i\, 0.21)~\pi G^{-1}$, with $G$ the Newton gravitation constant. The estimated uncertainty was of a 20\% from expected contribution at the next order in the unitarization process.

However, due to the null mass of the force carriers the PWAs in the case of gravity or Coulomb scattering are ill defined, being affected by  infrared (IR) divergences. The treatment offered in Ref.~\cite{Blas:2020och}, discussed in more detailed in Ref.~\cite{Blas:2020dyg}, was recently criticized by Ref.~\cite{Dobado:2022}, which also denied the existence of the graviball. The Ref.~\cite{Dobado:2022} introduces a completely different way of dealing with IR divergences, to find the needed perturbative PWAs in order to apply the unitarization method.

It is the aim of the present study to compare these two procedures for unitarizing infinite-range forces with the exactly solvable  non-relativistic Coulomb scattering. For that, we perturbatively calculate the Coulomb PWAs up to the one-loop level  in Sec.~\ref{sec.220708.1}, and unitarize them with both methods. It will turn out that while the unitarization method of Refs.~\cite{Blas:2020och,Blas:2020dyg} is meaningful and works as expected,  Sec.~\ref{sec.220709.1}, the newest one of Ref.~\cite{Dobado:2022} drives to mistaken PWAs and pole content,  Sec.~\ref{sec.220710.1}. A final discussion is given in Sec.~\ref{sec.220711.1}.

\section{Coulomb scattering up to one-loop (once-iterated Born series)}
\label{sec.220708.1}

Let us consider the calculation of the scattering of an electron in an external Yukawa potential
\begin{align}
  \label{220707.1}
  V(r)= Z\alpha\frac{e^{-\lambda r}}{r}
\end{align}
that in the limit $\lambda\to 0$ turns into Coulomb scattering, with $\alpha=e^2/4\pi$ the fine-structure constant.

 The Fourier transform of the Yukawa potential 
reads
\begin{align}
  \label{220707.2}
V(\vq)&=Z\alpha  \int d^3r\frac{e^{-i\vq \vr} e^{-\lambda r}}{r}=\frac{Ze^2}{\vq^2+\lambda^2}
\end{align}
with
\begin{align}
  \label{220707.3}
  \vq^2=(\vp'-\vp)^2=2\vp^2(1-\cos\theta)=4\vp^2\sin^2\frac{\theta}{2}~.
\end{align}
As usual $\theta$ is the scattering angle between the final $\vp'$ and initial $\vp$ momenta. In the following we take $Z=1$ for simplicity.\footnote{In our convention an attractive potential is positive. E.g. by taking $m$ as the electron mass we have the scattering of an electron by an attractive external Coulomb potential in the limit $\lambda\to 0$.} 

Dalitz provided in his seminal paper \cite{Dalitz:1951} the calculation of the scattering amplitude up to third order in the Born series.
His result is  given in the equation (3.4) of Ref.~\cite{Dalitz:1951} which, once multiplied by the factor $4\pi^3/m$ to conform with our normalization, becomes 
\begin{align}
  \label{220707.4}
  f(\vp')&=\frac{e^2}{4p^2\sin^2\frac{1}{2}\theta}
  \left\{
  1
  -i\frac{m\alpha}{p}
  \left(\log\sin^2\frac{1}{2}\theta+\log\frac{4p^2}{\lambda^2}\right)+\left(\frac{m\alpha}{p}\right)^2
  \left[-\frac{3}{4}
    \left(\log\frac{4p^2}{\lambda^2}\right)^2
    +\left[\log(\alpha\lambda)\right]^2+{\cal O}(\alpha^3)
    \right]
  \right\}~.
\end{align}
This formula keeps only the leading $\lambda$-dependence in the limit $\lambda\to 0$.\footnote{Let us notice that the exact Coulomb scattering amplitude is not affected by IR divergences because it is defined with respect to asymptotic Coulomb wave functions, and not plane waves. However, the latter are used as asymptotic states for $f(\vp')$ in Eq.~\eqref{220707.4}, and this is why it is IR divergent.}

Now, we work out the partial-wave projected $S$ wave up to the one-loop level.
The leading-order (LO) contribution is given by the angular average of $V(\vq)$ in Eq.~\eqref{220707.2},
\begin{align}
  \label{220708.1a}
  F_0^{(1)}(p)&=\frac{1}{2}\int_{-1}^{+1}d\cos\theta\frac{e^2}{2p^2(1-\cos\theta)+\lambda^2}=\frac{e^2}{2p^2}\log\frac{2p}{\lambda}~.
  \end{align}
The expression for the scattering amplitude from the once-iterated Born term, $T^{(2)}(\vp',\vp)$, is
\begin{align}
  \label{220708.1}
  T^{(2)}(\vp',\vp)&=\frac{me^4}{4\pi^3}\int\frac{d^3q}{[\lambda^2+(\vp'-\vq)^2][\lambda^2+(\vp-\vq)^2][q^2-p^2-i\ep]}~.
\end{align}
From this expression it is rather straightforward to take its $S$-wave projection, $F^{(2)}_0(p)$.\footnote{Note that Eq.~\eqref{220707.4} is not suitable to project in PWAs.} We then have
\begin{align}
  \label{220709.1}
  F^{(2)}_0(p)&= \frac{me^4}{16\pi^4}\int d\hat\vp' \int\frac{d^3q}{[\lambda^2+(\vp'-\vq)^2][\lambda^2+(\vp-\vq)^2][q^2-p^2-i\ep]}\\
  &=\frac{me^4}{16\pi^4}\int_0^\infty \frac{q^2 dq}{q^2-p^2-i\ep}\int \frac{d\hat \vq}{\lambda^2+(\vp-\vq)^2}
  \int \frac{d\hat\vp'}{\lambda^2+(\vp'-\vq)^2}=\frac{me^4}{16\pi^4}\int_0^\infty \frac{q^2 dq}{q^2-p^2-i\ep}\left[
    \int \frac{d\hat \vp'}{\lambda^2+(\vp'-\vq)^2}\right]^2\,.\nn
\end{align}
One could also have arrived to this result by considering the Lippmann-Schwinger equation in PWAs.

The imaginary part of $F_0^{(2)}(p)$ is restricted by perturbative unitarity in partial waves, which implies
\begin{align}
  \label{220709.2}
  \Im F_0^{(2)}(p)=\frac{mp}{2\pi}{F_0^{(1)}}^2=\frac{m e^2}{8\pi p^3}\left(\log\frac{2p}{\lambda}\right)^2~.
\end{align}
We can explicitly check that Eq.~\eqref{220709.1} for $F^{(2)}_0(p)$ satisfies it by direct calculation
\begin{align}
  \label{220709.3}
\Im F_0^{(2)}(p)&=\frac{me^4}{16\pi^4}\int_0^\infty dq q^2\pi \delta(q^2-p^2)\left[
  \int \frac{d\hat \vp'}{\lambda^2+(\vp'-\vp)^2}\right]^2
=\frac{m p e^4}{32\pi^3}\left[
  \int \frac{d\hat \vp'}{\lambda^2+(\vp'-\vp)^2}\right]^2~.
\end{align}
Now, by taking into account that $F_0^{(1)}(p)$ in Eq.~\eqref{220708.1a} is also given by
\begin{align}
  \label{220709.4}
  F_0^{(1)}(p)&=\frac{e^2}{4\pi}\int\frac{d\hat\vp'}{\lambda^2+(\vp'-\vp)^2}=\frac{e^2}{2p^2}\log\frac{2p}{\lambda}~,
\end{align}
the equality between Eqs.~\eqref{220709.2} and \eqref{220709.3} follows.

The real part of $F_0^{(2)}(p)$, given by the Cauchy principal value of the integral over $q$ in Eq.~\eqref{220709.1}, indeed vanishes for $\lambda\to 0$. Namely,
\begin{align}
\label{220709.5}
\Re F_0^{(2)}(p)&
=\lim_{\lambda\to 0}\frac{me^4}{16\pi^2 p^2}
\dashint_0^\infty \frac{dq}{q^2-p^2}\left[ \log\frac{\lambda^2+(p+q)^2}{\lambda^2+(p-q)^2}\right]^2=0~.
\end{align}
This is reflected in  Eq.~\eqref{220707.4}, which ${\cal O}(e^4)$ contribution is purely imaginary for real $p$.

We have calculated $F_0^{(2)}(p)$ numerically for $\lambda=0$, and checked both Eqs.~\eqref{220709.2} and \eqref{220709.5} for the real and imaginary parts of $F_0^{(2)}$, respectively .
From a numerical point of view we have found advantageous to take the limit $\lambda\to 0$ in the formula
\begin{align}
\label{220709.6}
\lim_{\lambda\to 0} \frac{me^4}{16\pi^2}\int_0^\infty \frac{dq}{q^2-p^2-i\lambda}\left[ \log\frac{\lambda^2+(p+q)^2}{\lambda^2+(p-q)^2}\right]^2~,
\end{align}
which shares the same  $\lambda\to 0$ limit as the original $F_2^{(0)}(p)$ in Eq.~\eqref{220709.5}. 

In summary, as a consequence of Eqs.~\eqref{220709.2} and \eqref{220709.5},
we have found that the $S$-wave projection of $F_0^{(2)}(p)$
is dominated with arbitrary precision for $\lambda\to 0$ by 
\begin{align}
\label{220709.7}
F_0^{(2)}(p)&=i\frac{m e^4}{8\pi p^3}\left(\log\frac{2p}{\lambda}\right)^2~,
\end{align}
Then, the $S$-wave PWA for (non-relativistic) Coulomb scattering  up to  one-loop (once-iterated Born series) is 
\begin{align}
  \label{220709.8}
F_0(p)&=F_0^{(1)}(p)+F_0^{(2)}(p)+{\cal O}(\alpha^3)=\frac{e^2}{2p^2}\log\frac{2p}{\lambda}+i\frac{m e^4}{8\pi p^3}\left(\log\frac{2p}{\lambda}\right)^2+{\cal O}(\alpha^3)~.
  \end{align}

Due to the infinite-range character of the electromagnetic interactions the PWAs are ill defined, and this is why $F_0(p)$ diverges for $\lambda\to 0$.
In the next section,  we discuss two unitarization methods for treating infinite-range interactions that have appeared recently in the literature.
Our method, the earliest \cite{Blas:2020och,Blas:2020dyg}, and the one of Ref.~\cite{Dobado:2022}, which has just appeared in the  literature. In the following we call {\bf method A} the former and {\bf method B} the latter.

\section{Application of the method A to Coulomb scattering}
\label{sec.220709.1}

The unitarization formula for the non-relativistic scattering is
\begin{align}
  \label{220709.9}
T_J(p)&=\left[V_J(p)^{-1}-i\frac{mp}{2\pi}\right]^{-1}~.
\end{align}
The interaction kernel $V_J(p)$ is expressed as a power-series expansion $V_J=V_J^{(1)}+V_J^{(2)}+{\cal O}(\alpha^3)$, and then matched with the perturbative power-series expansion of $T_J$ itself in order to determine the different $V_J^{(n)}$. The previous formula takes into account that
$\Im T_J^{-1}=- mp/2\pi$, and then one performs a dispersion relation along the RC
  \begin{align}
  \label{220709.10}
T_J^{-1}(p)&=V_J^{-1}-\frac{mp^2}{2\pi^2}\int \frac{dq^2 q}{(q^2-p^2-i\ep)q^2}
  \end{align}
  where the crossed-channel dynamics is included in $V_J^{-1}$. The result of the integration is simply $-imp/2\pi$, and Eq.~\eqref{220709.9} results.\footnote{One could think of including a subtraction constant together with the dispersive integral in Eq.~\eqref{220709.10}. However, this constant is reabsorbed in $V_J^{-1}$ in Eq.~\eqref{220709.9}. The Ref.~\cite{Blas:2020dyg} elaborates further on the suitability of taking this subtraction constant equal to zero based on the Sugawara-Kanazawa theorem, to which the interested reader is referred.}  
 When relativistic kinematics is used in the unitarity loop function, which would be expressed by an analogous dispersion relation as in Eq.~\eqref{220709.10},  also has a real part, see e.g. Eq.~(8.3) of Ref.~\cite{oller.book}. 

 We have already calculated perturbatively the $S$-wave PWA latter in Sec.~\ref{sec.220708.1}, though they are actually ill defined because of the IR divergences. It is at this point where we make use of the redefinition of the $S$ matrix by introducing the phase factor derived by Weinberg by resumming the exchanges of soft gravitons between interacting lines \cite{Weinberg:1965nx}, following Refs.~\cite{Blas:2020dyg,Blas:2020och}.  This phase factor, that we call $S_c$, for the  non-relativistic scattering of a charged particle by a Coulomb potential is
 \begin{align}
  \label{220709.11}
   S_c&=e^{2i\gamma \log\frac{\cal L}{\lambda}}~,~~\gamma=\frac{me^2}{p}~,
 \end{align}
and it was  firstly conjectured by Dalitz \cite{Dalitz:1959},

In Eq.~\eqref{220709.11} it is clear that the scale ${\cal L}$ is proportional to $2p$ since the resummed exchange of soft photons of $\lambda$ (which acts as an infrared regulator) implies a left-hand cut (LC) starting at $4p^2=-\lambda^2$. This is precisely the onset of the LC in the $\log {\cal L}/\lambda$ present in the exponent of $S_c$ for
 \begin{align}
  \label{220709.24}
   {\cal L}=\frac{2p}{a}~,
 \end{align}
 with the constant $a$ being necessarily independent of $p$.

 We then redefine the $S$ matrix by multiplying the standard $S$ matrix, $\bar{S}$, by $S_c^{-1}$, as introduced in Refs.~\cite{Blas:2020och,Blas:2020dyg}.
 The new $S$ matrix, $S$, 
 \begin{align}
  \label{220709.12a}
   S&=\bar{S}S_c^{-1}~.
 \end{align}
 is  free from IR divergences for the scattering problem with a potential $V(r)$, see the last section of Ref.~\cite{Weinberg:1965nx}.
  Regarding the value of $a$, it is important to stress that Eq.~\eqref{220709.11} is actually a resummation of the soft exchanges of photons under the assumption that their momenta are clearly smaller than the momentum $p$ of the external particles.
 However, as a resummation one could use it for convenience  with finite values of $a$ as well, with the pertinent extra contributions accounted for by the dependence of $S$ on this parameter. 
 From this perspective, one could then naturally consider  $a$ as a number larger than 1 but with a finite value. Indeed, it only enters logarithmically in the problem and we will take  in the following that $\log a={\cal O}(1)$. Later, we give the value of $a$ by comparing with the known exact Coulomb $S$ matrix in partial-wave amplitudes, as similarly done in Ref.~\cite{Blas:2020dyg}. We advance that in that case one has $\log a=\gamma_E+{\cal O}(\alpha^2)$. 

 It is worth stressing two facts regarding the previous equation for $S$. First, the new $S$ matrix $S$ is also unitary since $S_c$ is just a phase factor of unit modulus. Second, $S_c$ is independent of angle so that one can directly take the partial-wave projection on both sides of Eq.~\eqref{220709.12a} with the results that
 \begin{align}
  \label{220709.12}
   S_J&=\bar{S}_JS_c^{-1}~,
 \end{align}
 with $J$ the total angular momentum. 

 Let us work out the perturbative calculation of $S_J$ from Eq.~\eqref{220709.12}, which also teaches us  about the structure of the infrared divergences in $F_0^{(n)}$.
 These results can be  explicitly confirmed by the knowledge of $F_0$ up to one-loop level worked out in Sec.~\ref{sec.220708.1}.
 Working up to and including ${\cal O}(\alpha^2)$ contributions,
 \begin{align}
  \label{220709.13}
  S_J&=\left\{1+i\frac{mp}{\pi}(F_J^{(1)}+F_J^{(2)})\right\}\left\{1-2i\gamma\log\frac{{\cal L}}{\lambda}-2\gamma^2(\log\frac{\cl}{\lm})^2\right\}+{\cal O}(\alpha^3)\\
  &=
  1-2i\gamma\log\frac{\cl}{\lm}-2\gamma^2(\log\frac{\cl}{\lm})^2
  +i\frac{mp}{\pi}\left\{F_J^{(1)}\left[1-2i\gamma\log\frac{\cl}{\lm}\right]+F_J^{(2)}\right\}+{\cal O}(\alpha^3)~.\nn
 \end{align}
 Considering this expression at ${\cal O}(\alpha)$, 
 and since $S_J$ is free of IR divergences, 
 if follows that the dependence of $F_J^{(1)}$ on $\log\lambda$ is $-\frac{e^2}{2p^2}\log\lambda$ 
  in agreement with Eq.~\eqref{220708.1a} for $J=0$. As shown in Ref.~\cite{Blas:2020dyg} the IR divergent contribution in $F_J^{(1)}$ is the same for all the PWAs and is removed by that from the Weinberg phase at ${\cal O}(\alpha)$.
 
 Considering now Eq.~\eqref{220709.13} at ${\cal O}(\alpha^2)$ it requires that the IR divergences entering at this order must affect only the imaginary part $F_J^{(2)}$, since $F_J^{(1)}$ is real as it is the partial-wave projection of the Born term with angular momentum $J$. This is manifest in Eq.~\eqref{220709.7} for $J=0$. 

 Since $S_J$ is unitary we define an associated $T$ matrix free of IR divergences by the standard relation between $S$ and $T$ matrices in PWAs,
 \begin{align}
  \label{220709.16}
S_J(p)=1+i\frac{mp}{\pi}T_J(p)~.
 \end{align}
 We then read from Eq.~\eqref{220709.13} the following expression up to ${\cal O}(\alpha^2)$ for $T_J^{(n)}(p)$.
 \begin{align}
  \label{220709.17}
T_J^{(1)}(p)&=F_J^{(1)}(p)-\frac{e^2}{2p^2}\log\frac{\cl}{\lm}~,\\
\label{220709.18a}
T_J^{(2)}(p)&=F_J^{(2)}(p)-i F_J^{(1)}(p)\frac{me^2}{2\pi p}\log\frac{\cl}{\lm}
+i\frac{me^4}{8\pi p^3}
\left(\log\frac{\cl}{\lm}\right)^2~.
 \end{align}

 For the case $J=0$ we can use the calculated expressions for $F_0^{(1)}$ and $F_0^{(2)}$ above and obtain
 \begin{align}
   \label{220709.18}
   T_0^{(1)}(p)&=\frac{e^2}{2p^2}\log\frac{2p}{\cl}=\frac{e^2}{2p^2}\log a~,\\
   \label{220709.19}
   T_0^{(2)}(p)&=i\frac{m e^4}{8\pi p^3}\left(\log\frac{2p}{\cl}\right)^2=i\frac{m e^4}{8\pi p^3}\left(\log a\right)^2~.
 \end{align}
 Now, in order to determine $V_0^{(1)}$ and $V_0^{(2)}$ we match the general unitarization formula of Eq.~\eqref{220709.9}
 with $T_0^{(1)}+T_0^{(2)}$,
 \begin{align}
\label{220709.20}
T_0^{(1)}+T_0^{(2)}=\left[\frac{1}{V_0^{(1)}+V_0^{(2)}}-i\frac{mp}{2\pi}\right]^{-1}=V_0^{(1)}+ V_0^{(2)}+i\frac{mp}{2\pi}{V_0^{(1)}}^2+{\cal O}(\alpha^3)~.
   \end{align}
 Therefore,
\begin{align}
  \label{220709.21}
  V_0^{(1)}(p)&=T_0^{(1)}(p)=\frac{e^2}{2p^2}\log a~,\\
  \label{220709.22}
  V_0^{(2)}(p)&=0~.
\end{align} 
Thus, up to ${\cal O}(\alpha^2)$ in the expansion of the interaction kernel $V_0(p)$ our expression for $T_0(p)$ is
\begin{align}
  \label{220709.23}
T_0(p)&=\left(\frac{2p^2}{e^2\log a}-i\frac{mp}{2\pi}\right)^{-1}~.
\end{align}

At this point, we compare it with the exact partial-wave decomposition of the on-shell Coulomb scattering \cite{Kang:1962}
\begin{align}
  \label{220710.1a}
T(\vp',\vp)&=\frac{\pi}{imp}\sum_J (2J+1)\left(\frac{\Gamma(J+1-i \gamma)}{\Gamma(J+1+i \gamma)}-1\right)P_J(\cos\theta)~,
\end{align}
so that
\begin{align}
  \label{220710.1}
  S_J(p)&=\frac{\Gamma(J+1-i \gamma)}{\Gamma(J+1+i \gamma)}=e^{2i\sigma_\ell}~,\\
  \label{220710.2}
  T_J(p)&=\frac{\pi}{i m p}\left(S_J(p)-1\right)~,
\end{align}
with $\sigma_\ell$ the pure Coulomb phase shifts. By comparing Eqs.~\eqref{220709.9} and \eqref{220710.2} we can determine the exact expression for $V_J(p)$ to all orders
\begin{align}
  \label{220710.3}
  V_J(p)&=-i\frac{2\pi}{m p}\frac{\Gamma(1+J-i\gamma)-\Gamma(1+J+i\gamma)}{\Gamma(1+J-i\gamma)+\Gamma(1+J+i\gamma)}~.
\end{align}
Its perturbative expansion gives
\begin{align}
  \label{220710.4}
V_J(p)&=-\frac{e^2}{2p^2}\psi_0(1+J)+{\cal O}(\alpha^3)~,
\end{align}
where $\psi_n(z)=d^{n+1}\log\Gamma(z)/dz^{n+1}$. In particular for $J=0$ we have that\footnote{There is an explicit formula for $\psi_0(n)=-\gamma+\theta(n-1)\sum_{k=1}^{n-1}1/k$\,.}

\begin{align}
  \label{220710.5}
  V_0(p)&=\frac{e^2}{2p^2}\gamma_E+{\cal O}(\alpha^3)~.
\end{align}
By comparison with Eq.~\eqref{220709.21}, this implies the following LO expression for $\log a$
\begin{align}
  \label{220710.6}
  \log a&=\gamma_E+{\cal O}(\alpha^2)~.
\end{align}

Contrarily to the statements in Ref.~\cite{Dobado:2022}, where the method B is developed, this precise relation for $\log a$ clearly shows that the constant $a$ is not a cutoff.
It has a precise value that has been fixed by matching with the exact solution.
However,  a cutoff is an auxiliary scale that is always taken within a convenient range of values or sent to infinity (as is typically  done in renormalizable quantum field theories).\footnote{The Eq.~\eqref{220710.6} can also be obtained by  employing a screened Coulomb potential and the known asymptotic behavior of the radial Coulomb wave functions, without actually solving Coulomb scattering to get the exact $S_J(p)$  \cite{Blas:2020dyg}.}

Then, we write our final unitarized expression for $T_0$ obtained by determining $V_0(p)$ up to ${\cal O}(\alpha^2)$ included,
\begin{align}
  \label{220710.7}
T_0(p)&=\frac{2\pi}{mp}\left(\frac{1}{\gamma_E\gamma}-i\right)^{-1}~.
\end{align}
The position of the bound state is then given by
\begin{align}
  \label{220710.8a}
  p^{(1)}=i\,m \alpha \gamma_E~,
\end{align}
to be compared with the exact result
\begin{align}
  \label{220710.8}
  p_{\rm exact}=i\,m \alpha~,
\end{align}
so that there is an extra factor $\gamma_E\approx 0.58$, with an error at LO of a 40\%.
We should notice that our thumb of rule estimate for $\log a={\cal O}(1)$, {\it i.e.} taking $\log a=1$, would give the exact value for the binding momentum.

One can indeed consider higher orders in the expansion in powers of $\alpha$ of $V_0(p)$
by using its exact expression in Eq.~\eqref{220710.4}. In this way, we also obtain that $V_0^{(2)}(p)=0$, and indeed only odd powers in the expansion are non zero. The third order contribution $V_0^{(3)}$ is
\begin{align}
  \label{220710.9}
V_0^{(3)}=\frac{2\pi}{m p}\frac{\gamma^3}{3}(\gamma_E-\zeta(3))~,
\end{align}
with $\zeta(z)$ the Riemann zeta function.
By taking $V_0=\sum_{i=1}^3V_0^{(i)}$ one obtains
\begin{align}
  \label{220710.10}
T_0=\frac{2\pi}{mp}\left(\left\{\gamma_E \gamma+\frac{\gamma_E-\zeta(3)}{3}\gamma^3\right\}^{-1}-i\right)^{-1}~.
\end{align}
The pole position for the bound state in the $p$-axis is 
\begin{align}
  p^{(3)}&=0.950\, p_{\rm exact}~,
\end{align}
and the uncertainty has reduced to a 5\% only. The inclusion of still higher orders decreases further the uncertainty \cite{Blas:2020dyg}. This is explicitly worked out in Table~\ref{tab.110711.1} where the binding momentum is given with respect to the exact solution.   The convergence with increasing order is  remarkably fast.

     \begin{table}
       \begin{center}
         \begin{tabular}{|l|llll|}
           \hline
           $n$ & 1 & 3 & 5 & 7 \\
           \hline
           ${p^{(n)}}/p_{\rm exact}$ & $0.58$ & $0.95$ & $1.00$ & $1.00$  \\
\hline
         \end{tabular}
         \caption{Pole positions in the complex $p$-plane corresponding to the deepest bound state from $T_0$, Eq.~\ref{220709.9}, with $V_0(p)=\sum_{i=1}^nV_0^{(i)}(p)$. 
           The pole position is given with respect to the exact pole position $im\alpha$.
           \label{tab.110711.1}}
       \end{center}
       \end{table}

\begin{figure}[ht]
\begin{center}       \includegraphics[width=.6\textwidth,angle=0]{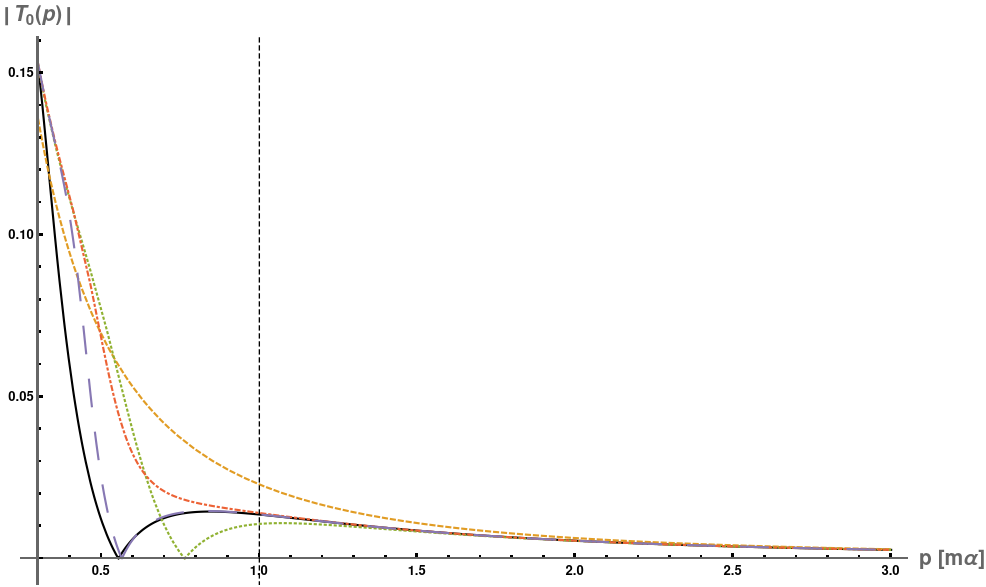}
  \caption{The modulus of the Coulomb  $S$-wave PWA given by the exact result (black solid line) is compared with those obtained by applying the unitarization formula Eq.~\eqref{220709.9} up to order ${\cal O}(\alpha^n)$ in the calculation of $V_0=\sum_{i=1}^n V^{(i)}_0(p)$.
    The lines with $n=1,$ 3, 5 and 7 are given by the orange dashed, green dotted, red dash-dotted and  magenta long-dashed lines, respectively. In the figure energy units are taken such that $m\alpha=1$, with the vertical line indicating this value for $p$. \label{fig.220711.1}}
\end{center}     \end{figure}

     The good behavior of the approach by increasing the precision up to which $V_0(p)$ is calculated is also reflected by comparing directly the PWA $T_0(p)$ calculated from the unitarization formula, Eq.~\eqref{220709.9}, with the exact one in
     Eq.~\eqref{220710.2}. This is shown in Fig.~\ref{fig.220711.1} where the absolute value of $T_0(p)$ is shown for the exact solution, and the LO and NLO unitarized expression by the solid, dashed and dot-dashed lines, respectively. We also show the resulting unitarized $T_0(p)$ up to $n=7$, with a steady improvement in the reproduction of the exact $S$-wave PWA. Due to the essential singularity of the exact Coulomb PWA at $p=0$ its reproduction for $p\ll m\alpha$ requires to include many higher orders and is not practical. As noticed in Ref.~\cite{Blas:2020dyg} Coulomb scattering becomes trivial for $p\to\infty$, so that its reproduction with just a few orders of $V_0$ is very good for $p\gtrsim m\alpha$ (this value of $p$ is indicated by the dashed vertical line in the figure). 

\section{Application of the method B to Coulomb scattering}
\label{sec.220710.1}

Recently, Ref.~\cite{Dobado:2022} has performed a criticism to the results of Ref.~\cite{Blas:2020dyg} advocating that no resonance appears in graviton-graviton scattering in the energy on which Einstein gravity could be treated as a quantum effective field theory. 
The Ref.~\cite{Dobado:2022} does not take into account the Weinberg phase but directly consider the perturbative calculation of the PWAs in terms of a graviton mass $\lambda$ which is finally sent to zero. The unitarization formula employed by Ref.~\cite{Dobado:2022} is the Inverse Amplitude Method (IAM) to one loop \cite{Roiesnel:1980gd,Truong:1988zp,Dobado:1989qm}. We refer to Ref.~\cite{Oller:2020guq} for a recent review on unitarization methods with historical remarks. 

Since we know the perturbative PWAs $F_0^{(1)}$ and $F_0^{(2)}$ (i.e. up to one-loop order),  we are now in position to apply the unitarization method B of Ref.~\cite{Dobado:2022} to non-relativistic Coulomb scattering and compare its results with the known exact solution. In particular, we discuss the movement of the position of the poles as $\lambda\to 0$, according to the prescriptions of Ref.~\cite{Dobado:2022}.

The one-loop IAM unitarization formula reads
\begin{align}
  \label{220711.1}
T_0(p)&=\frac{F_0^{(1)}(p)^2}{F_0^{(1)}(p)-F_0^{(2)}(p)}~.
\end{align}
We introduce the variable $y=p/m\alpha$ and take units such that $2m\alpha=1$, so that the IAM $T_0(p)$ adopts the simpler form 
\begin{align}
  \label{220711.2}
T_0(p)&=\frac{8\pi\alpha\log(y/\lambda)}{y\left[y-i\log(y/\lambda)\right]}~.
\end{align}

We can then look for poles in this expression for  $y\neq 0$. First, it is important to notice that
for $\lambda>0$ there is no bound state pole (despite Coulomb scattering is not resonant and only has bound states). This is clear if we rewrite $y=i\kappa$, with $\kappa>0$, so that the secular equation $i\kappa+\pi/2-i\log\kappa/\lambda=0$ results, which clearly has no solution for real and positive $\kappa$.

The poles of $T_0(p)$ for $y\neq 0$ correspond to the solutions of the equation
\begin{align}
  \label{220711.3}
  y=i\log(y/\lambda)~.
\end{align}
We find values of $y$ with $\Im p<0$ that solve this equation and that correspond to resonance poles.   
For small values of $\lambda$ (such that $\lambda\ll -\log \lambda$) the approximate solution is given by $p=\lambda-i\delta$, with $\delta/\lambda$ vanishing as $\lambda\to 0$.  This can be seen by direct substitution in Eq.~\eqref{220711.3}, where the error done is ${\cal O}(\lambda)$ which is much smaller than $-\log\lambda$ which drives the equation, and this is why it is a good approximation for $\lambda\ll 1$. For instance, for $\lambda=0.1$ we have that $-(\log\lambda)/\lambda>23$. We have calculated numerically the solution for some values of $\lambda$, as shown in Table~\ref{tab.110711.2}.
As a result, one cannot reproduce any bound state characteristic of the Hydrogen atom by employing the IAM formula Eq.~\eqref{220711.2} for small $\lambda$. When $\lambda\to 0$ the resonance pole tends asymptotically to $p\to 0$, while the exact solution is $im\alpha$ and, of course, it remains finite and fixed (as dictated by experiment, or even by the anthropic principle). 

     \begin{table}
       \begin{center}
         \begin{tabular}{|l|llll|}
           \hline
           $\lambda~ [2m\alpha]$ & $0.5$ & $10^{-1}$ & $10^{-2}$ & $10^{-4}$ \\
           \hline
        $p/\lambda$ & $0.79-i\,0.32$ & $0.99-i\,0.10$ & $1.00-i0.01$ & $1.00-i0.00$  \\
\hline
         \end{tabular}
         \caption{Pole positions in the complex $p$-plane corresponding to resonances arising from the method of Ref.~\cite{Dobado:2022}, Eq.~\eqref{220711.1}, for decreasing values of $\lambda$ (with values given in units of $2m\alpha$). Notice that the pole position asymptotically tends to $\lambda$ and hence vanishes. In contrast, the exact pole position is $i m\alpha$.
           \label{tab.110711.2}}
       \end{center}
       \end{table}

\begin{figure}[ht]
\begin{center}       \includegraphics[width=.6\textwidth,angle=0]{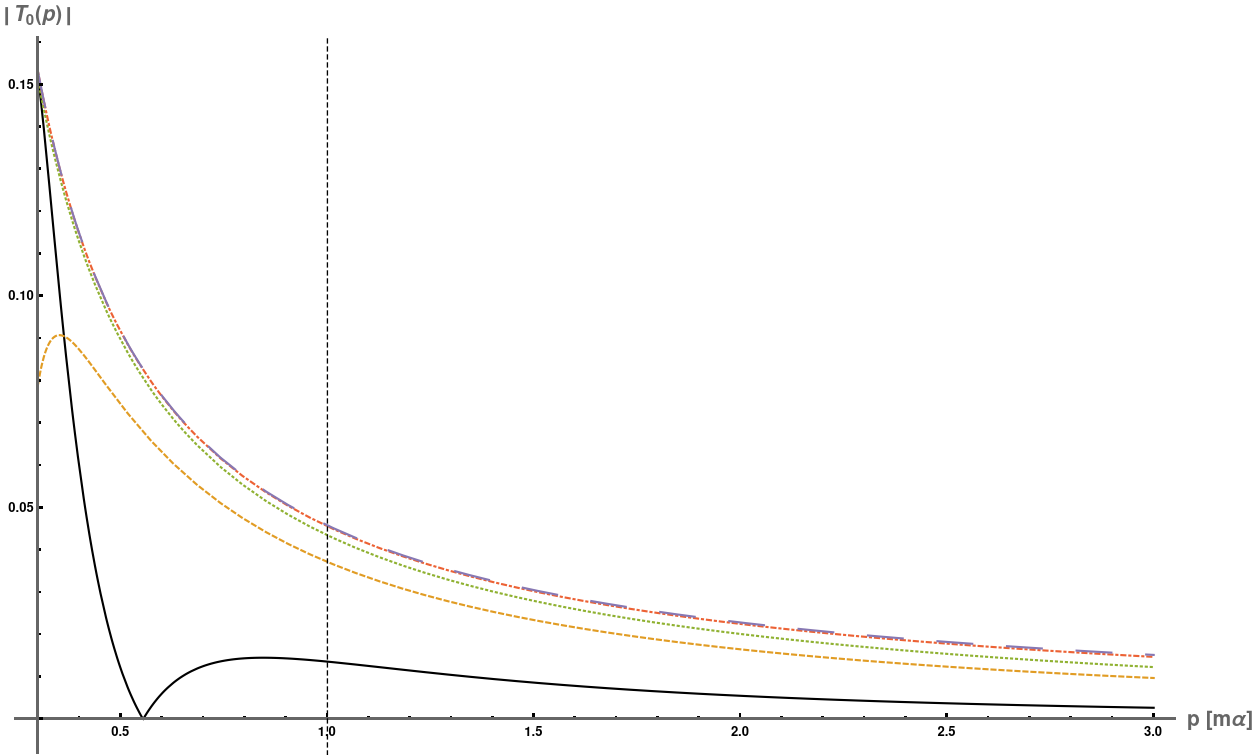}
\caption{The modulus of the Coulomb  $S$-wave PWA given by the exact result (solid line) is compared with those obtained by applying the IAM-method of Ref.~\cite{Dobado:2022}, Eq.~\eqref{220711.1} for $\lambda=0.5$ (orange dashed), $10^{-1}$ (green dotted),  $10^{-4}$ (red dot-dashed) and $10^{-12}$ (magenta long-dashed line). In the figure energy units are taken such that $m\alpha=1$, with the vertical line indicating this value for $p$. \label{fig.220711.2}}
\end{center}     \end{figure}

     The gross failure of the approach of Ref.~\cite{Dobado:2022}  for unitarizing long-range forces in order to account for Coulomb scattering is also manifest if we compare $T_0(p)$, given by Eq.~\eqref{220711.1},  with the known exact Coulomb $S$-wave PWA.
     This is shown in  Fig.~\ref{fig.220711.2} where the modulus of the exact Coulomb $S$-wave (solid line) is compared with $T_0(p)$ obtained from the IAM with values of $\lambda=1/2$ (dashed), $10^{-1}$ (dotted),  $10^{-4}$ (dot-dashed) and $10^{-12}$ (long-dashed line). The last two lines cannot be distinguished. As in Fig.~\ref{fig.220711.1} the unit of energy is $m\alpha=1$.  We observe that the Coulomb $S$-wave is not reproduced either quantitatively or qualitatively speaking. As a matter of fact, for $\lambda\to 0$ and finite $p$ ($p\neq 0$ nor $p\to \infty$)  the PWA $T_0(p)$ from Eq.~\eqref{220711.1} degenerates into $-i2\pi/mp$ without any dynamical content. 

     Indeed, there is a fundamental difference between the exact solution and the one provided by the method B of Ref.~\cite{Dobado:2022}. While the former has no LC the latter has one for $p<0$ (due to the dependence on $\log 2p/\lambda$ with $\lambda\to 0$), so that the analytical properties of the PWA resulting from the method B are, strictly speaking, incorrect. Regarding this, we recall that $\Gamma(z)$ is a meromorphic function in the complex $z$ plane, having only simple poles for $z=-n$ with $n=0$ or in $\mathbb{N}$. In this way, the exact $S_J(p)$ from Eq.~\eqref{220710.1}  has only RC.

     \section{Discussion}
     \label{sec.220711.1}
     In summary, the study of non-relativistic Coulomb scattering is a clear counterexample, which shows that the method of Ref.~\cite{Dobado:2022} for unitarizing forces of infinite range is not suitable.
     Its failure lies in the bad treatment of IR divergences at the level of the scattering amplitudes.
 The authors of Ref.~\cite{Dobado:2022}  keep  in the perturbative PWAs the infrared regulator $\lambda$, interpreted  as a photon (graviton) mass,    so that they take the limit $\lambda\to 0$, in order not to contradict the experimental limit for its mass (e.g. of the order of $10^{-30}$~eV for the graviton \cite{pdg}).  It is also worth noticing that the reason for the wrong PWA $T_0$ provided by the method of Ref.~\cite{Dobado:2022} is not due to the use of the particular unitarization method IAM, Eq.~\eqref{220711.1}, but because of the inadequate treatment  of the IR divergences. In this respect, if the IAM formula is applied with the perturbative IR-safe PWAs $T_0^{(1)}$ and $T_0^{(2)}$, replacing $F_0^{(1)}$ and $F_0^{(2)}$, respectively, one indeed obtains the same result as by using the unitarization method A \cite{Blas:2020dyg,Blas:2020och} because
     \begin{align}
       \label{220711.4}
       \frac{{T_0^{(1)}}^2}{T_0^{(1)}-T_0^{(2)}}=
     \frac{ {T_0^{(1)}}^2}{T_0^{(1)}-i\frac{mp}{2\pi}{T_0^{(1)}}^2}=\frac{1}{ \frac{1}{T_0^{(1)}}-i\frac{mp}{2\pi}}~.
\end{align}
       
However, the presence of $\lambda$ is removed once the Weinberg's phase is used, so it is  replaced by  ${\cal L}=2p/a$ with the parameter $a$ being  in principle calculable within the scattering theory under study, like in our case with Coulomb scattering.     In this way, the dependence on $p$ of the IR-safe perturbative PWAs changes, so that, instead of $F_0^{(1,2)}(p)$, one now has $T_0^{(1,2)}(p)$. 

 Thus, in $F_0^{(1,2)}(p)$  the logarithmic diverging contribution $\log 2p/\lambda$ for $\lambda\to 0$ and finite $p$ is replaced by the constant $\log a=\gamma_E={\cal O}(1)$. 
In this way, we have shown that we are able to properly reproduce within the unitarization method of Eq.~\eqref{220709.9} the non-relativistic Coulomb scattering amplitude, with its right analytical properties, improve the accuracy of its reproduction by including higher orders, and adequately accounting for the pole position of the ground state.

We would like to end with some extra discussions to answer some other criticisms of Ref.~\cite{Dobado:2022} on \cite{Blas:2020och,Blas:2020dyg}. The dependence of the graviball pole on the scale $\Lambda$ introduced in Ref.~\cite{Blas:2020och} is only logarithmic, while Ref.~\cite{Dobado:2022}  in its Appendix I mistakenly states that it is quadratic. The confusion arises because $\Lambda$  is taken as the unit of energy when giving the graviball $s_P$  in Ref.~\cite{Blas:2020och} as $s_P=(0.07-i\,0.21)\,\Lambda^2$. 
     
Another unfortunate statement in the same Appendix I of Ref.~\cite{Dobado:2022} concerns the claimed inability of the unitarization method of  Refs.~\cite{Blas:2020och,Blas:2020dyg} to generate the $\rho(770)$ in hadron physics. As explained long time ago in Ref.~\cite{nd}, due to the KSFR relation \cite{Kawarabayashi:1966kd,Riazuddin:1966sw}, the $\rho(770)$ can be generated by adjusting adequately the subtraction constant $a_{\pi\pi}$ in the unitarity loop function for $\pi\pi$ scattering in Eq.~(5.17) of Ref.~\cite{Blas:2020dyg}. Denoting by $T_1^{\pi\pi}(s)$ the  $\pi\pi$ $P$-wave scattering amplitude in the chiral limit, with massless pions,
     \begin{align}
       \label{220714.1}
       T_1^{\pi\pi}(s)=\left(\frac{6f^2}{s}+\frac{1}{16\pi^2}\left\{a_{\pi\pi}+\log\frac{-s}{\Lambda^2}\right\}\right)^{-1}~.
     \end{align}
Let us notice that the first term is nothing else but the inverse of the LO $\pi\pi$ $P$-wave amplitude (instead of the scalar isoscalar counterpart employed in Refs.~\cite{Blas:2020dyg,Blas:2020och} for discussing the $\sigma$ resonance. The chiral limit is used for analogy with the massless character of gravitons and photons, similarly as in Refs.~\cite{Blas:2020dyg,Blas:2020och}).  
We first take the scale $\Lambda=M_\rho$, the mass of $\rho(770)$, which is a natural value for the hadronic non-perturbative chiral expansion scale. The value
     \begin{align}
       \frac{a_{\pi\pi}}{(4\pi)^2}=-\frac{6f^2}{M_\rho^2}\approx 0.086~,
     \end{align}
was obtained in Ref.~\cite{nd} taking into account the KSFR relation. With this result Eq.~\eqref{220714.1} has a $\rho$-resonance pole at $s_\rho=(0.56-i\,0.14)$~GeV$^2$, which is very close indeed to its physical value $s_\rho\approx (0.58-i\,0.12)$~GeV$^2$ \cite{pdg}. The pole position $s_\rho$ only depends logarithmically on the scale $\Lambda$, so that for $\Lambda=4\pi f_\pi$ \cite{Georgi:1984}(another typical value for the chiral expansion scale, around $1.5 M_\rho$) we find $s_\rho=(0.53-i\,0.12)$~GeV$^2$.

\appendix



\section*{Acknowledgements}

The author is grateful to J. Mart\'{\i}n-Camalich and D.~Blas for many discussions along the years on related issues, and for reading the manuscript.
I also acknowledge interesting discussions with J.~R.~Pel\'aez.   
This work has been supported in part by the MICINN
AEI (Spain) Grant No. PID2019–106080GB-C22/AEI/
10.13039/501100011033, and by the EU Horizon 2020
research and innovation programme, STRONG-2020
project, under Grant agreement No. 824093.

\bibliographystyle{unsrt}
\bibliography{ref}

\end{document}